\documentclass[conference]{IEEEtran}
\IEEEoverridecommandlockouts
\usepackage{cite}
\usepackage{amsmath,amssymb,amsfonts}
\usepackage{algorithmic}
\usepackage{graphicx}
\usepackage{textcomp}
\usepackage{xcolor}
\usepackage{comment}

\usepackage{multicol}
\usepackage{subfig}

\def\BibTeX{{\rm B\kern-.05em{\sc i\kern-.025em b}\kern-.08em
    T\kern-.1667em\lower.7ex\hbox{E}\kern-.125emX}}

\newtheorem{proposition}{Proposition}

\newtheorem{corollary}{Corollary}
\newtheorem{lemma}{Lemma}

\usepackage{amssymb}
\usepackage{amsfonts}
\usepackage{mathrsfs}
\usepackage{xspace}
\usepackage{bm}
\usepackage{upgreek}

\newcommand{\safemath}[2]{\newcommand{#1}{\ensuremath{#2}\xspace}}

\safemath{\bma}{\mathbf{a}}
\safemath{\bmb}{\mathbf{b}}
\safemath{\bmc}{\mathbf{c}}
\safemath{\bmd}{\mathbf{d}}
\safemath{\bme}{\mathbf{e}}
\safemath{\bmf}{\mathbf{f}}
\safemath{\bmg}{\mathbf{g}}
\safemath{\bmh}{\mathbf{h}}
\safemath{\bmi}{\mathbf{i}}
\safemath{\bmj}{\mathbf{j}}
\safemath{\bmk}{\mathbf{k}}
\safemath{\bml}{\mathbf{l}}
\safemath{\bmm}{\mathbf{m}}
\safemath{\bmn}{\mathbf{n}}
\safemath{\bmo}{\mathbf{o}}
\safemath{\bmp}{\mathbf{p}}
\safemath{\bmq}{\mathbf{q}}
\safemath{\bmr}{\mathbf{r}}
\safemath{\bms}{\mathbf{s}}
\safemath{\bmt}{\mathbf{t}}
\safemath{\bmu}{\mathbf{u}}
\safemath{\bmv}{\mathbf{v}}
\safemath{\bmw}{\mathbf{w}}
\safemath{\bmx}{\mathbf{x}}
\safemath{\bmy}{\mathbf{y}}
\safemath{\bmz}{\mathbf{z}}
\safemath{\bmzero}{\mathbf{0}}
\safemath{\bmone}{\mathbf{1}}

\bmdefine{\biad}{a}
\bmdefine{\bibd}{b}
\bmdefine{\bicd}{c}
\bmdefine{\bidd}{d}
\bmdefine{\bied}{e}
\bmdefine{\bifd}{f}
\bmdefine{\bigd}{g}
\bmdefine{\bihd}{h}
\bmdefine{\biid}{i}
\bmdefine{\bijd}{j}
\bmdefine{\bikd}{k}
\bmdefine{\bild}{l}
\bmdefine{\bimd}{m}
\bmdefine{\bind}{n}
\bmdefine{\biod}{o}
\bmdefine{\bipd}{p}
\bmdefine{\biqd}{q}
\bmdefine{\bird}{r}
\bmdefine{\bisd}{s}
\bmdefine{\bitd}{t}
\bmdefine{\biud}{u}
\bmdefine{\bivd}{v}
\bmdefine{\biwd}{w}
\bmdefine{\bixd}{x}
\bmdefine{\biyd}{y}
\bmdefine{\bizd}{z}

\bmdefine{\bixid}{\xi}
\bmdefine{\bilambdad}{\lambda}
\bmdefine{\bimud}{\mu}
\bmdefine{\bithetad}{\theta}
\bmdefine{\biphid}{\phi}
\bmdefine{\bideltad}{\delta}

\safemath{\bmia}{\biad}
\safemath{\bmib}{\bibd}
\safemath{\bmic}{\bicd}
\safemath{\bmid}{\bidd}
\safemath{\bmie}{\bied}
\safemath{\bmif}{\bifd}
\safemath{\bmig}{\bigd}
\safemath{\bmih}{\bihd}
\safemath{\bmii}{\biid}
\safemath{\bmij}{\bijd}
\safemath{\bmik}{\bikd}
\safemath{\bmil}{\bild}
\safemath{\bmim}{\bimd}
\safemath{\bmin}{\bind}
\safemath{\bmio}{\biod}
\safemath{\bmip}{\bipd}
\safemath{\bmiq}{\biqd}
\safemath{\bmir}{\bird}
\safemath{\bmis}{\bisd}
\safemath{\bmit}{\bitd}
\safemath{\bmiu}{\biud}
\safemath{\bmiv}{\bivd}
\safemath{\bmiw}{\biwd}
\safemath{\bmix}{\bixd}
\safemath{\bmiy}{\biyd}
\safemath{\bmiz}{\bizd}

\safemath{\bmxi}{\bixid}
\safemath{\bmlambda}{\bilambdad}
\safemath{\bmmu}{\bimud}
\safemath{\bmtheta}{\bithetad}
\safemath{\bmphi}{\biphid}
\safemath{\bmdelta}{\bideltad}

\safemath{\bA}{\mathbf{A}}
\safemath{\bB}{\mathbf{B}}
\safemath{\bC}{\mathbf{C}}
\safemath{\bD}{\mathbf{D}}
\safemath{\bE}{\mathbf{E}}
\safemath{\bF}{\mathbf{F}}
\safemath{\bG}{\mathbf{G}}
\safemath{\bH}{\mathbf{H}}
\safemath{\bI}{\mathbf{I}}
\safemath{\bJ}{\mathbf{J}}
\safemath{\bK}{\mathbf{K}}
\safemath{\bL}{\mathbf{L}}
\safemath{\bM}{\mathbf{M}}
\safemath{\bN}{\mathbf{N}}
\safemath{\bO}{\mathbf{O}}
\safemath{\bP}{\mathbf{P}}
\safemath{\bQ}{\mathbf{Q}}
\safemath{\bR}{\mathbf{R}}
\safemath{\bS}{\mathbf{S}}
\safemath{\bT}{\mathbf{T}}
\safemath{\bU}{\mathbf{U}}
\safemath{\bV}{\mathbf{V}}
\safemath{\bW}{\mathbf{W}}
\safemath{\bX}{\mathbf{X}}
\safemath{\bY}{\mathbf{Y}}
\safemath{\bZ}{\mathbf{Z}}

\safemath{\bZero}{\mathbf{0}}
\safemath{\bOne}{\mathbf{1}}
\safemath{\bDelta}{\mathbf{\Delta}}
\safemath{\bLambda}{\mathbf{\UpLambda}}
\safemath{\bPhi}{\mathbf{\Upphi}}
\safemath{\bSigma}{\mathbf{\Upsigma}}
\safemath{\bOmega}{\mathbf{\Upomega}}
\safemath{\bTheta}{\mathbf{\Uptheta}}

\bmdefine{\biAd}{A}
\bmdefine{\biBd}{B}
\bmdefine{\biCd}{C}
\bmdefine{\biDd}{D}
\bmdefine{\biEd}{E}
\bmdefine{\biFd}{F}
\bmdefine{\biGd}{G}
\bmdefine{\biHd}{H}
\bmdefine{\biId}{I}
\bmdefine{\biJd}{J}
\bmdefine{\biKd}{K}
\bmdefine{\biLd}{L}
\bmdefine{\biMd}{M}
\bmdefine{\biOd}{N}
\bmdefine{\biPd}{O}
\bmdefine{\biQd}{P}
\bmdefine{\biRd}{R}
\bmdefine{\biSd}{S}
\bmdefine{\biTd}{T}
\bmdefine{\biUd}{U}
\bmdefine{\biVd}{V}
\bmdefine{\biWd}{W}
\bmdefine{\biXd}{X}
\bmdefine{\biYd}{Y}
\bmdefine{\biZd}{Z}

\bmdefine{\biDelta}{\Delta}
\bmdefine{\biLambda}{\Lambda}
\bmdefine{\biPhi}{\Phi}
\bmdefine{\biSigma}{\Sigma}
\bmdefine{\biOmega}{\Omega}
\bmdefine{\biTheta}{\Theta}

\safemath{\bimA}{\biAd}
\safemath{\bimB}{\biBd}
\safemath{\bimC}{\biCd}
\safemath{\bimD}{\biDd}
\safemath{\bimE}{\biEd}
\safemath{\bimF}{\biFd}
\safemath{\bimG}{\biGd}
\safemath{\bimH}{\biHd}
\safemath{\bimI}{\biId}
\safemath{\bimJ}{\biJd}
\safemath{\bimK}{\biKd}
\safemath{\bimL}{\biLd}
\safemath{\bimM}{\biMd}
\safemath{\bimN}{\biNd}
\safemath{\bimO}{\biOd}
\safemath{\bimP}{\biPd}
\safemath{\bimQ}{\biQd}
\safemath{\bimR}{\biRd}
\safemath{\bimS}{\biSd}
\safemath{\bimT}{\biTd}
\safemath{\bimU}{\biUd}
\safemath{\bimV}{\biVd}
\safemath{\bimW}{\biWd}
\safemath{\bimX}{\biXd}
\safemath{\bimY}{\biYd}
\safemath{\bimZ}{\biZd}

\safemath{\bimDelta}{\biDelta}
\safemath{\bimLambda}{\biLambda}
\safemath{\bimPhi}{\biPhi}
\safemath{\bimSigma}{\biSigma}
\safemath{\bimOmega}{\biOmega}
\safemath{\bimTheta}{\biTheta}

\safemath{\setA}{\mathcal{A}}
\safemath{\setB}{\mathcal{B}}
\safemath{\setC}{\mathcal{C}}
\safemath{\setD}{\mathcal{D}}
\safemath{\setE}{\mathcal{E}}
\safemath{\setF}{\mathcal{F}}
\safemath{\setG}{\mathcal{G}}
\safemath{\setH}{\mathcal{H}}
\safemath{\setI}{\mathcal{I}}
\safemath{\setJ}{\mathcal{J}}
\safemath{\setK}{\mathcal{K}}
\safemath{\setL}{\mathcal{L}}
\safemath{\setM}{\mathcal{M}}
\safemath{\setN}{\mathcal{N}}
\safemath{\setO}{\mathcal{O}}
\safemath{\setP}{\mathcal{P}}
\safemath{\setQ}{\mathcal{Q}}
\safemath{\setR}{\mathcal{R}}
\safemath{\setS}{\mathcal{S}}
\safemath{\setT}{\mathcal{T}}
\safemath{\setU}{\mathcal{U}}
\safemath{\setV}{\mathcal{V}}
\safemath{\setW}{\mathcal{W}}
\safemath{\setX}{\mathcal{X}}
\safemath{\setY}{\mathcal{Y}}
\safemath{\setZ}{\mathcal{Z}}
\safemath{\emptySet}{\varnothing}

\safemath{\colA}{\mathscr{A}}
\safemath{\colB}{\mathscr{B}}
\safemath{\colC}{\mathscr{C}}
\safemath{\colD}{\mathscr{D}}
\safemath{\colE}{\mathscr{E}}
\safemath{\colF}{\mathscr{F}}
\safemath{\colG}{\mathscr{G}}
\safemath{\colH}{\mathscr{H}}
\safemath{\colI}{\mathscr{I}}
\safemath{\colJ}{\mathscr{J}}
\safemath{\colK}{\mathscr{K}}
\safemath{\colL}{\mathscr{L}}
\safemath{\colM}{\mathscr{M}}
\safemath{\colN}{\mathscr{N}}
\safemath{\colO}{\mathscr{O}}
\safemath{\colP}{\mathscr{P}}
\safemath{\colQ}{\mathscr{Q}}
\safemath{\colR}{\mathscr{R}}
\safemath{\colS}{\mathscr{S}}
\safemath{\colT}{\mathscr{T}}
\safemath{\colU}{\mathscr{U}}
\safemath{\colV}{\mathscr{V}}
\safemath{\colW}{\mathscr{W}}
\safemath{\colX}{\mathscr{X}}
\safemath{\colY}{\mathscr{Y}}
\safemath{\colZ}{\mathscr{Z}}

\safemath{\opA}{\mathbb{A}}
\safemath{\opB}{\mathbb{B}}
\safemath{\opC}{\mathbb{C}}
\safemath{\opD}{\mathbb{D}}
\safemath{\opE}{\mathbb{E}}
\safemath{\opF}{\mathbb{F}}
\safemath{\opG}{\mathbb{G}}
\safemath{\opH}{\mathbb{H}}
\safemath{\opI}{\mathbb{I}}
\safemath{\opJ}{\mathbb{J}}
\safemath{\opK}{\mathbb{K}}
\safemath{\opL}{\mathbb{L}}
\safemath{\opM}{\mathbb{M}}
\safemath{\opN}{\mathbb{N}}
\safemath{\opO}{\mathbb{O}}
\safemath{\opP}{\mathbb{P}}
\safemath{\opQ}{\mathbb{Q}}
\safemath{\opR}{\mathbb{R}}
\safemath{\opS}{\mathbb{S}}
\safemath{\opT}{\mathbb{T}}
\safemath{\opU}{\mathbb{U}}
\safemath{\opV}{\mathbb{V}}
\safemath{\opW}{\mathbb{W}}
\safemath{\opX}{\mathbb{X}}
\safemath{\opY}{\mathbb{Y}}
\safemath{\opZ}{\mathbb{Z}}
\safemath{\opZero}{\mathbb{O}}
\safemath{\identityop}{\opI}

\safemath{\veca}{\bma}
\safemath{\vecb}{\bmb}
\safemath{\vecc}{\bmc}
\safemath{\vecd}{\bmd}
\safemath{\vece}{\bme}
\safemath{\vecf}{\bmf}
\safemath{\vecg}{\bmg}
\safemath{\vech}{\bmh}
\safemath{\veci}{\bmi}
\safemath{\vecj}{\bmj}
\safemath{\veck}{\bmk}
\safemath{\vecl}{\bml}
\safemath{\vecm}{\bmm}
\safemath{\vecn}{\bmn}
\safemath{\veco}{\bmo}
\safemath{\vecp}{\bmp}
\safemath{\vecq}{\bmq}
\safemath{\vecr}{\bmr}
\safemath{\vecs}{\bms}
\safemath{\vect}{\bmt}
\safemath{\vecu}{\bmu}
\safemath{\vecv}{\bmv}
\safemath{\vecw}{\bmw}
\safemath{\vecx}{\bmx}
\safemath{\vecy}{\bmy}
\safemath{\vecz}{\bmz}

\safemath{\veczero}{\bmzero}
\safemath{\vecone}{\bmone}
\safemath{\vecxi}{\bmxi}
\safemath{\veclambda}{\bmlambda}
\safemath{\vecmu}{\bmmu}
\safemath{\vectheta}{\bmtheta}
\safemath{\vecphi}{\bmphi}
\safemath{\vecdelta}{\bmdelta}

\safemath{\matA}{\bA}
\safemath{\matB}{\bB}
\safemath{\matC}{\bC}
\safemath{\matD}{\bD}
\safemath{\matE}{\bE}
\safemath{\matF}{\bF}
\safemath{\matG}{\bG}
\safemath{\matH}{\bH}
\safemath{\matI}{\bI}
\safemath{\matJ}{\bJ}
\safemath{\matK}{\bK}
\safemath{\matL}{\bL}
\safemath{\matM}{\bM}
\safemath{\matN}{\bN}
\safemath{\matO}{\bO}
\safemath{\matP}{\bP}
\safemath{\matQ}{\bQ}
\safemath{\matR}{\bR}
\safemath{\matS}{\bS}
\safemath{\matT}{\bT}
\safemath{\matU}{\bU}
\safemath{\matV}{\bV}
\safemath{\matW}{\bW}
\safemath{\matX}{\bX}
\safemath{\matY}{\bY}
\safemath{\matZ}{\bZ}
\safemath{\matzero}{\bmzero}

\safemath{\matDelta}{\bDelta}
\safemath{\matLambda}{\bLambda}
\safemath{\matPhi}{\bPhi}
\safemath{\matSigma}{\bSigma}
\safemath{\matOmega}{\bOmega}
\safemath{\matTheta}{\bTheta}

\safemath{\matidentity}{\matI}
\safemath{\matone}{\matO}

\safemath{\rnda}{A}
\safemath{\rndb}{B}
\safemath{\rndc}{C}
\safemath{\rndd}{D}
\safemath{\rnde}{E}
\safemath{\rndf}{F}
\safemath{\rndg}{G}
\safemath{\rndh}{H}
\safemath{\rndi}{I}
\safemath{\rndj}{J}
\safemath{\rndk}{K}
\safemath{\rndl}{L}
\safemath{\rndm}{M}
\safemath{\rndn}{N}
\safemath{\rndo}{O}
\safemath{\rndp}{P}
\safemath{\rndq}{Q}
\safemath{\rndr}{R}
\safemath{\rnds}{S}
\safemath{\rndt}{T}
\safemath{\rndu}{U}
\safemath{\rndv}{V}
\safemath{\rndw}{W}
\safemath{\rndx}{X}
\safemath{\rndy}{Y}
\safemath{\rndz}{Z}

\safemath{\rveca}{\bimA}
\safemath{\rvecb}{\bimB}
\safemath{\rvecc}{\bimC}
\safemath{\rvecd}{\bimD}
\safemath{\rvece}{\bimE}
\safemath{\rvecf}{\bimF}
\safemath{\rvecg}{\bimG}
\safemath{\rvech}{\bimH}
\safemath{\rveci}{\bimI}
\safemath{\rvecj}{\bimJ}
\safemath{\rveck}{\bimK}
\safemath{\rvecl}{\bimL}
\safemath{\rvecm}{\bimM}
\safemath{\rvecn}{\bimN}
\safemath{\rveco}{\bomO}
\safemath{\rvecp}{\bimP}
\safemath{\rvecq}{\bimQ}
\safemath{\rvecr}{\bimR}
\safemath{\rvecs}{\bimS}
\safemath{\rvect}{\bimT}
\safemath{\rvecu}{\bimU}
\safemath{\rvecv}{\bimV}
\safemath{\rvecw}{\bimW}
\safemath{\rvecx}{\bimX}
\safemath{\rvecy}{\bimY}
\safemath{\rvecz}{\bimZ}

\safemath{\rvecxi}{\bmxi}
\safemath{\rveclambda}{\bmlambda}
\safemath{\rvecmu}{\bmmu}
\safemath{\rvectheta}{\bmtheta}
\safemath{\rvecphi}{\bmphi}

\safemath{\rmatA}{\bimA}
\safemath{\rmatB}{\bimB}
\safemath{\rmatC}{\bimC}
\safemath{\rmatD}{\bimD}
\safemath{\rmatE}{\bimE}
\safemath{\rmatF}{\bimF}
\safemath{\rmatG}{\bimG}
\safemath{\rmatH}{\bimH}
\safemath{\rmatI}{\bimI}
\safemath{\rmatJ}{\bimJ}
\safemath{\rmatK}{\bimK}
\safemath{\rmatL}{\bimL}
\safemath{\rmatM}{\bimM}
\safemath{\rmatN}{\bimN}
\safemath{\rmatO}{\bimO}
\safemath{\rmatP}{\bimP}
\safemath{\rmatQ}{\bimQ}
\safemath{\rmatR}{\bimR}
\safemath{\rmatS}{\bimS}
\safemath{\rmatT}{\bimT}
\safemath{\rmatU}{\bimU}
\safemath{\rmatV}{\bimV}
\safemath{\rmatW}{\bimW}
\safemath{\rmatX}{\bimX}
\safemath{\rmatY}{\bimY}
\safemath{\rmatZ}{\bimZ}

\safemath{\rmatDelta}{\bimDelta}
\safemath{\rmatLambda}{\bimLambda}
\safemath{\rmatPhi}{\bimPhi}
\safemath{\rmatSigma}{\bimSigma}
\safemath{\rmatOmega}{\bimOmega}
\safemath{\rmatTheta}{\bimTheta}

\usepackage{amssymb}
\usepackage{amsfonts}
\usepackage{mathrsfs}
\usepackage{xspace}
\usepackage{bm}
\usepackage{fancyref}
\usepackage{textcomp}

\usepackage{multirow}
\usepackage{stmaryrd}

\newenvironment{textbmatrix}{	\setlength{\arraycolsep}{2.5pt}%
								\big[\begin{matrix}}{\end{matrix}\big]%
								\raisebox{0.08ex}{\vphantom{M}}}

\def\be{\begin{equation}}
\def\ee{\end{equation}}
\def\een{\nonumber \end{equation}}
\def\mat{\begin{bmatrix}}
\def\emat{\end{bmatrix}}
\def\btm{\begin{textbmatrix}}
\def\etm{\end{textbmatrix}}

\def\ba#1\ea{\begin{align}#1\end{align}}
\def\bas#1\eas{\begin{align*}#1\end{align*}}
\def\bs#1\es{\begin{split}#1\end{split}} 
\def\bg#1\eg{\begin{gather}#1\end{gather}}
\def\bml#1\eml{\begin{multline}#1\end{multline}}
\def\bi#1\ei{\begin{itemize}#1\end{itemize}}

\safemath{\dirac}{\delta}					
\safemath{\krond}{\dirac}					

\safemath{\upto}{\uparrow}
\safemath{\downto}{\downarrow}
\safemath{\iu}{j}							
\safemath{\ev}{\lambda}						
\safemath{\hilseqspace}{l^{2}}				
\newcommand{\banachfunspace}[1]{\setL^{#1}}	
\safemath{\hilfunspace}{\banachfunspace{2}}	

\safemath{\SNR}{\textsf{SNR}} 				
\safemath{\PAR}{\textsf{PAR}} 				
\safemath{\No}{N_0}							
\safemath{\Es}{E_s}							
\safemath{\Eb}{E_b}							
\safemath{\EbNo}{\frac{\Eb}{\No}}
\safemath{\EsNo}{\frac{\Es}{\No}}

\DeclareMathOperator{\CHop}{\ensuremath{\opH}} 
\safemath{\tvir}{\rndh_{\CHop}}				
\safemath{\tvtf}{\rndl_{\CHop}}				
\safemath{\spf}{\rnds_{\CHop}}				
\safemath{\bff}{H_{\CHop}}					

\safemath{\ircf}{r_{h}}						
\safemath{\tftvcf}{r_{s}}					
\safemath{\tfcf}{r_{l}}						
\safemath{\bfcf}{r_{H}}						

\safemath{\tcorr}{c_h}						
\safemath{\scf}{c_{s}}						
\safemath{\tfcorr}{c_{l}}					
\safemath{\fcorr}{c_{H}}						

\safemath{\mi}{I}							
\safemath{\capacity}{C}						

\safemath{\normal}{\mathcal{N}}			
\safemath{\jpg}{\mathcal{CN}}			
\safemath{\mchain}{\leftrightarrow}		

\safemath{\dB}{\,\mathrm{dB}}
\safemath{\dBm}{\,\mathrm{dBm}}
\safemath{\Hz}{\,\mathrm{Hz}}
\safemath{\kHz}{\,\mathrm{kHz}}
\safemath{\MHz}{\,\mathrm{MHz}}
\safemath{\GHz}{\,\mathrm{GHz}}
\safemath{\s}{\,\mathrm{s}}
\safemath{\ms}{\,\mathrm{ms}}
\safemath{\mus}{\,\mathrm{\text{\textmu}s}}
\safemath{\ns}{\,\mathrm{ns}}
\safemath{\ps}{\,\mathrm{ps}}
\safemath{\meter}{\,\mathrm{m}}
\safemath{\mm}{\,\mathrm{mm}}
\safemath{\cm}{\,\mathrm{cm}}
\safemath{\m}{\,\mathrm{m}}
\safemath{\W}{\,\mathrm{W}}
\safemath{\mW}{\, \mathrm{mW}}
\safemath{\J}{\,\mathrm{J}}
\safemath{\K}{\,\mathrm{K}}
\safemath{\bit}{\,\mathrm{bit}}
\safemath{\nat}{\,\mathrm{nat}}

\safemath{\define}{\triangleq}			

\safemath{\equivalent}{\sim}
\safemath{\distas}{\sim}					
\safemath{\sdiff}{\Delta}				

\safemath{\reals}{\mathbb{R}}
\safemath{\positivereals}{\reals_{+}}
\safemath{\integers}{\mathbb{Z}}
\safemath{\posint}{\integers_{+}}
\safemath{\naturals}{\mathbb{N}}
\safemath{\posnaturals}{\naturals_{+}}
\safemath{\complexset}{\mathbb{C}}
\safemath{\rationals}{\mathbb{Q}}

\newcommand*{\fancyrefapplabelprefix}{app}		
\newcommand*{\fancyrefthmlabelprefix}{thm}		
\newcommand*{\fancyreflemlabelprefix}{lem}		
\newcommand*{\fancyrefcorlabelprefix}{cor}		
\newcommand*{\fancyrefdeflabelprefix}{def}		
\newcommand*{\fancyrefproplabelprefix}{prop}	
\newcommand*{\fancyrefobslabelprefix}{obs}		
\newcommand*{\fancyrefalglabelprefix}{alg}		
\newcommand*{\fancyrefasmlabelprefix}{asm}	    

\frefformat{vario}{\fancyrefseclabelprefix}{Section~#1}
\frefformat{vario}{\fancyrefthmlabelprefix}{Theorem~#1}
\frefformat{vario}{\fancyreflemlabelprefix}{Lemma~#1}
\frefformat{vario}{\fancyrefcorlabelprefix}{Corollary~#1}
\frefformat{vario}{\fancyrefdeflabelprefix}{Definition~#1}
\frefformat{vario}{\fancyrefobslabelprefix}{Observation~#1}
\frefformat{vario}{\fancyrefasmlabelprefix}{Assumption~#1}
\frefformat{vario}{\fancyreffiglabelprefix}{Fig.~#1}
\frefformat{vario}{\fancyrefapplabelprefix}{Appendix~#1} 
\frefformat{vario}{\fancyrefproplabelprefix}{Proposition~#1}
\frefformat{vario}{\fancyrefalglabelprefix}{Algorithm~#1}
\frefformat{vario}{\fancyrefeqlabelprefix}{(#1)}

\safemath{\dictab}{[\,\dicta\,\,\dictb\,]}

\safemath{\ysig}{\bmy}
\safemath{\ysighat}{\hat{\ysig}}
\safemath{\ysigdim}{M}
\safemath{\xsig}{\bmx}
\safemath{\xsigdim}{N}
\safemath{\nx}{n_x}
\safemath{\zsig}{\bmz}
\safemath{\zsigdim}{\ysigdim}
\safemath{\rsig}{\bmr}
\safemath{\Adict}{\bA}
\safemath{\Adicttilde}{\widetilde{\Adict}}
\safemath{\Adictdim}{\outputdim\times\xsigdim}
\safemath{\avec}{\bma}
\safemath{\avectilde}{\tilde{\avec}}
\safemath{\Bdict}{\bB}
\safemath{\Bdicttilde}{\widetilde{\Bdict}}
\safemath{\Cdict}{\bC}
\safemath{\cvec}{\bmc}
\safemath{\Ddict}{\bD}
\safemath{\Ddictdim}{\ysigdim\times\xsigdim}
\safemath{\dvec}{\bmd}
\safemath{\Ddicttilde}{\widetilde{\bD}}
\safemath{\Bonb}{\bB}
\safemath{\bvec}{\bmb}
\safemath{\Bonbdim}{\ysigdim\times\ysigdim}
\safemath{\noise}{\bmn}
\safemath{\noisedim}{\ysigim}
\safemath{\err}{\bme}
\safemath{\errdim}{\ysigdim}
\safemath{\errset}{\setE}
\safemath{\nerr}{n_e}
\safemath{\delop}{\bP_\errset}
\safemath{\delopc}{\bP_{{\errset}^c}}

\safemath{\cplxi}{\imath}
\safemath{\cplxj}{\jmath}

\safemath{\dict}{\matD}
\safemath{\inputdim}{N}		
\safemath{\outputdim}{M}		
\safemath{\sparsity}{S}	
\safemath{\inputdimA}{{N_a}}	
\safemath{\inputdimB}{{N_b}}	
\safemath{\elemA}{{n_a}}	
\safemath{\elemB}{{n_b}}	
\safemath{\resA}{\matR_a}	
\safemath{\resB}{\matR_b}	
\safemath{\subD}{\matS} 
\safemath{\subA}{\matS_a} 
\safemath{\subB}{\matS_b} 
\safemath{\dicta}{\matA} 	
\safemath{\dictb}{\matB} 	
\safemath{\hollowS}{H}
\safemath{\hollowA}{H_a}
\safemath{\hollowB}{H_b}
\safemath{\cross}{Z}
\safemath{\coh}{\mu_d}			
\safemath{\coha}{\mu_a}			
\safemath{\cohb}{\mu_b}			
\safemath{\mubs}{\nu}	
\safemath{\cohm}{\mu_m} 
\safemath{\dictset}{\setD}	
\safemath{\dictsetp}{\dictset(\coh,\coha,\cohb)}	
\safemath{\dictsetgen}{\dictset_\text{gen}}
\safemath{\dictsetgenp}{\dictsetgen(\coh)}
\safemath{\dictsetonb}{\dictset_\text{onb}}
\safemath{\dictsetonbp}{\dictsetonb(\coh)}

\safemath{\leftside}{U}
\safemath{\rightsideA}{R_a}
\safemath{\rightsideB}{R_b}

\safemath{\indexS}{\setI_S} 

\safemath{\na}{n_a}			
\safemath{\nb}{n_b}			
\safemath{\coeffa}{p_i}	
\safemath{\coeffb}{q_j}	
\safemath{\seta}{\setP}		
\safemath{\setb}{\setQ}     
\safemath{\setw}{\setW}	
\safemath{\setz}{\setZ}	
\safemath{\cola}{\veca}		
\safemath{\colb}{\vecb}		
\safemath{\cold}{\vecd}		
\safemath{\inputvec}{\vecx} 	
\safemath{\error}{\vece}	
\safemath{\noiseout}{\vecz} 	
\safemath{\inputvecel}{x}
\safemath{\inputveca}{\vecx_a}
\safemath{\inputvecb}{\vecx_b}
\safemath{\outputvec}{\vecy}	
\safemath{\lambdamin}{\lambda_{\mathrm{min}}}

\safemath{\elltwo}{\ell_2}
\safemath{\ellone}{\ell_1}
\safemath{\ellzero}{\ell_0}
\safemath{\ellinf}{\ell_\infty}
\safemath{\ellinftilde}{\ell_{\widetilde\infty}}
\safemath{\licard}{Z(\coh,\coha,\cohb)}
\safemath{\xsol}{\hat{x}}
\safemath{\xbord}{x_b}		
\safemath{\xstat}{x_s}		
\safemath{\xstatLone}{\tilde{x}_s}
\safemath{\order}{\mathcal{O}} 
\safemath{\scales}{\Theta} 
\safemath{\ones}{\mathbf{1}} 
\safemath{\zeroes}{\mathbf{0}} 
\safemath{\thlone}{\kappa(\coh,\cohb)} 
\safemath{\constoneA}{\delta} 
\safemath{\constoneB}{\epsilon} 
\safemath{\nlarge}{L}				   
\safemath{\sumlarge}{S_\nlarge}
\safemath{\maxlarger}{P_\nlarge}	   
\safemath{\Pzero}{\textrm{P0}}	
\safemath{\Pone}{\textrm{P1}}
\safemath{\vecfir}{\vecw}			 
\safemath{\vecsec}{\vecz}
\safemath{\elvecfir}{w}              
\safemath{\elvecsec}{z}				 
\safemath{\nlargefir}{n}
\safemath{\normout}{\gamma}
\safemath{\auxfun}{h}
\safemath{\supp}{\textrm{supp}}

\safemath{\indexa}{\ell}
\safemath{\indexb}{r}
\safemath{\indexc}{i}
\safemath{\indexd}{j}

\safemath{\project}{P}

\begin{document}

\title{Exploiting Mutual Coupling Characteristics for Channel Estimation in Holographic MIMO}


\author{\IEEEauthorblockN{
Nikolaos Kolomvakis and   
Emil Bj\"{o}rnson   
}                                     
Division of Communication Systems, KTH Royal Institute of Technology, Stockholm, Sweden\\ \thanks{This work was supported by the FFL18-0277 grant from the Swedish Foundation for Strategic Research.}
Email: \{nikkol, emilbjo\}@kth.se}

\maketitle

\begin{abstract}
Holographic multiple-input multiple-output (MIMO) systems represent a spatially constrained MIMO architecture with a massive number of antennas with small antenna spacing as a close approximation of a spatially continuous electromagnetic aperture. Accurate channel modeling is essential for realizing the full potential of this technology. In this paper, we investigate the impact of mutual coupling and spatial channel correlation on the estimation precision in holographic MIMO systems, as well as the importance of knowing their characteristics.
We demonstrate that neglecting mutual coupling can lead to significant performance degradation for the minimum mean squared error estimator, emphasizing its critical consideration when designing estimation algorithms. Conversely, the least-squares estimator is resilient to mutual coupling but only yields good performance in high signal-to-noise ratio regimes. Our findings provide insights into how to design efficient estimation algorithms in holographic MIMO systems, aiding its practical implementation.
\end{abstract}


\section{Introduction}

Massive MIMO antenna systems are key components of fifth-generation (5G) networks, playing a vital role in achieving high spectral efficiency across wireless channels through beamforming and spatial multiplexing. The spectral efficiency of massive MIMO grows monotonically with the number of antennas, so it would be desirable to have nearly infinitely many antennas. Therefore, in the next generation of wireless systems, we can expect the deployment of hundreds or even thousands of antennas at base stations (BSs). 
Ongoing research efforts are exploring the transition from moderately large antenna arrays in conventional massive MIMO to extremely large arrays, possibly spanning tens of meters. These systems, known as Extremely Large Aperture Arrays (ELAAs), promise significant performance improvements\cite{BJORNSON20193}.

However, practical constraints limit the number of half-wavelength-spaced antennas that can be deployed at conventional towers and rooftop locations. In particular, site owner restrictions, weight considerations, and wind load limits impose constraints on array dimensions.
If we keep adding more antennas into a fixed aperture, we converge to a spatially continuous aperture, called large intelligent surfaces or holographic MIMO\cite{Pizzo2022}.
In practice, we can approximate it with a discrete aperture with an antenna spacing much smaller than half-wavelength. 
While it might deliver holographic MIMO gains, dense antenna configurations inevitably lead to electromagnetic interactions and mutual coupling among antennas. 

Even if a holographic MIMO system has thousands of antennas, we can obtain channel state information (CSI) from a single pilot symbol if it is transmitted by the user device. The minimum mean squared error (MMSE) estimator gives optimal estimation accuracy but requires accurate knowledge of the channel covariance matrix,
 which is challenging to identify in practice.
To circumvent this issue, \cite{ChanEstimOzlem2022} proposed a channel estimator that exploits only the correlation characteristics imposed by the array geometry but not the ones imposed by the user-specific propagation channel.
A low-complexity method was proposed in \cite{LucaDFT2023} based on a discrete Fourier transform approximation.
There is also a body of work on beam training (e.g., \cite{Cui2022a}), which can be used in the special case of sparse channels. These previous works have neglected the mutual coupling effects.

In this paper, we aim to provide new insights into the impact of spatial channel correlation and mutual coupling on channel estimation in holographic MIMO. We consider several MMSE-type estimators and compare them with the classical least squares (LS) estimator. Specifically, when considering the Bayesian MMSE estimator, we distinguish between two kinds of prior information: mutual coupling characteristics and user-specific spatial channel correlation. We analyze the impact when the base station (BS) has access to one or both of them.

Our analytical and numerical results highlight the critical importance of considering mutual coupling when implementing channel estimation. When the MMSE estimator is used, ignoring mutual coupling can lead to significant performance degradation, 
while having access to the spatial channel correlation is less important.
Finally, our findings indicate that the LS estimator exhibits resilience to the absence of mutual coupling information but only performs well at high signal-to-noise ratio (SNR).

\section{System Model}

We consider the uplink of a single-cell holographic massive MIMO system where the base station (BS) serves single-antenna users. This paper analyzes the channel estimation procedures for one arbitrary user, assigned an orthogonal pilot. The BS is equipped with a uniform planar array (UPA) with $M$ elements, which are arranged in the $yz$-plane with $M_y$ and $M_z$ elements on the horizontal ($y$-axis) and vertical ($z$-axis) axes, respectively, such that $M=M_y\cdot M_z$. For notational convenience, the antenna elements are indexed row-by-row by $m\in \{1,2,...,M\}$, that is $m=m_yM_z+m_z$, for $m_y=1,...,M_y$ and $m_z=1,...,M_z$. Hence, the array response vector when a plane wave impinges on the UPA from the azimuth angle $\varphi$ and elevation angle $\theta$ is written as
\begin{align} \label{eq:array-response}
        \veca(\varphi,\theta) = \left[ e^{j\veck(\varphi,\theta)^T \vecr_1} ,\dots, e^{j\veck(\varphi,\theta)^T \vecr_M} \right]^T,
\end{align}
where $\veck(\varphi,\theta) \triangleq \frac{2\pi}{\lambda}\left(\cos\theta \cos\varphi, \cos\theta \sin\varphi,\sin\theta \right)^T$ is the wave vector and $\vecr_m$ is the location of the $m$-th antenna with respect to the origin of the Cartesian coordinate system. 

Assuming that the first element of the UPA is placed at the origin, we have $\vecr_m = \left( 0, r_y(m)d_y, r_z(m)d_z \right)^T$,
where $r_z(m) \triangleq \mod(m,M_z)$ and $r_y(m) \triangleq \lfloor \frac{m}{M_z} \rfloor$ are the horizontal and vertical indices of the $m$-th element, respectively, while $\mod$ is the modulo operator. Moreover, $d_y$ and $d_z$ are the horizontal and vertical inter-element spacing, respectively.

The channel between the BS and the considered user is denoted by the vector $\vech\in \mathbb{C}^M$, and we consider the general correlated Rayleigh fading distribution \cite{CorrRayEmil2021}, 
expressed as
\begin{align}\label{eq:rayleigh_channel}
    \vech = \matR^{1/2}\vech_{\tt iid},
\end{align} 
where the column vector $\vech_{\tt iid}\in \mathbb{C}^M$ contains independent and identically distributed (i.i.d.) zero-mean and unit-variance complex Gaussian entries. The matrix $\matR \triangleq \mathbb{E}\{\vech\vech^H\}\in \mathbb{C}^{M\times M}$ represents the spatial correlation of the channel $\vech$, where the $(n,m)$-th entry of $\matR$ is given by\cite{ChanEstimOzlem2022,CorrRayEmil2021}
\begin{align}\label{eq:spatial-correlation-formula}
    [\matR]_{n,m} = \iint \limits_{-\pi/2}^{-\pi/2} f(\varphi,\theta)
     e^{j\veck(\varphi,\theta)^T \left(\vecr_n - \vecr_m\right)}d\theta d\varphi.
\end{align}
Here, $f(\cdot)$ is the spatial scattering function, which describes the angular multipath distribution (including their relative strength) and the directivity gain of the antennas.

The double integral in \eqref{eq:spatial-correlation-formula} can be numerically computed for any spatial scattering function, and certain functions also yield closed-form expressions, as demonstrated in the subsequent section. A typical example, detailed in \cite{CorrRayEmil2021}, is an isotropic scattering environment, where multipath components exhibit equal strength in all directions, and antennas are isotropic.


\subsection{Spatial Correlation}

We proceed by analytically deriving the spatial correlation matrix under the assumption of an isotropic scattering environment and a UPA equipped with half-wavelength dipole antennas (in contrast to \cite{CorrRayEmil2021}). The directivity of such antennas can be approximated by $D = 1.67\cos^3(\theta)$ \cite{Balanis}, thus yielding the spatial scattering function as follows:
\begin{align}
    f(\varphi,\theta) = \frac{1.67}{2\pi}\cos^4(\theta),
\end{align}
where the additional $\cos$-term accounts for the spherical coordinate system\cite{Emil2024Book}.

\begin{proposition}\label{prop:spatial-correlation-isotropic}
    In the scenario of isotropic scattering in the half-space in front of the UPA, which is equipped with half-wavelength dipole antenna elements, the $(n,m)$-th entry of the spatial correlation matrix $\matR_{\tt iso}$ in \eqref{eq:spatial-correlation-formula} is given by
    \begin{align}
    \begin{split}\label{eq:isotropic-correlation-formula}
            [\matR_{\tt iso}]_{n,m} = \sum_{k=0}^{\infty} \sum_{l=0}^{k} \alpha_{k,l} \left( \left( r_z(n)-r_z(m)\right) d_z \right)^{2(k-l)}\\
            \times\left( \left( r_y(n)-r_y(m)\right) d_y \right)^{2l} ,       
    \end{split}
    \end{align}
    where the scalar value $\alpha_{k,l}$ is given by
    \begin{align}
    \begin{split}
        \alpha_{k,l} = (-1)^k \frac{1}{(2k)!} \binom{2k}{2l}\cdot \binom{2l}{l}\cdot \binom{2k-2l}{k-l}\cdot \pi^{2k+2}\\
        \times \frac{1.67}{2\pi}\frac{(2l+3)!!}{(2k+4)(2k+2)...(2k-2l+2)}   .  
    \end{split}
    \end{align}
    \begin{IEEEproof}
    See Appendix A.
    \end{IEEEproof}
\end{proposition}
The infinite summation in \eqref{eq:isotropic-correlation-formula} converges rapidly for small values of the inter-element spacing, so only a few terms are needed. This expression is computationally tractable and accurate for holographic MIMO, typically applicable to inter-element spacings spanning several carrier wavelengths.

For a more realistic scenario involving non-isotropic scattering environments, \cite{ChanEstimOzlem2022} derived the spatial correlation matrix considering scattered waves reaching the BS via a set of $N$ angular clusters, representing various objects within the environment.
The spatial scattering function for the $n$-th cluster, assuming dipole antennas, is given by \cite[Eq.~(10)]{ChanEstimOzlem2022}
\begin{align}
    f_n(\delta_n,\epsilon_n) = \mathcal{A}\cdot \mathcal{P}_n\cos^{4}(\theta_n+\epsilon_n) e^{\frac{\cos(2\delta_n)}{4\sigma^2_{\phi}}} e^{\frac{\cos(2\epsilon_n)}{4\sigma^2_{\theta}}},
\end{align}
where $\mathcal{P}_n\ge 0$ is the normalized power of cluster $n$ and the scalar $\mathcal{A}$ is selected such that the total power is unit. The $n$-th cluster is centered around the nominal azimuth and elevation angles $\varphi_n$ and $\theta_n$, respectively, while the variables $\delta_n$ and $\epsilon_n$ denote the respective angular deviations. From expression in \eqref{eq:spatial-correlation-formula}, the $(m,l)$-th entry of the non-isotropic spatial correlation matrix $\matR_{\tt clu}$ becomes
\begin{align}\label{eq:spatial-correlation-formula-clusters}
    \begin{split}
        [\matR_{\tt clu}]_{m,l} =\mathcal{A}\sum_{n=1}^{N} \mathcal{P}_n\int \limits_{-\theta_n-\pi/2}^{-\theta_n+\pi/2} \hspace{2mm} \int \limits_{-\varphi_n-\pi/2}^{-\varphi_n+\pi/2} f_n(\delta_n,\epsilon_n)\\
        \times e^{j\veck(\varphi_n+\delta_n,\theta_n+\epsilon_n)^T \left(\vecr_m - \vecr_l\right)}d\delta_n d\epsilon_n.
    \end{split}
\end{align}
This integral can be computed numerically, but it is computationally demanding for large arrays. A closed-form approximation that is tight for narrow angular clusters is provided in \cite[Lemma 2]{ChanEstimOzlem2022}.

It is worth mentioning that the span of the correlation matrix with isotropic scattering spans the entire angular domain as shown in\cite[Lemma 3]{ChanEstimOzlem2022}. Hence, for a given array geometry, the subspace spanned by the columns of $\matR_{\tt iso}$ contains the subspace spanned by any other $\matR$, such as $\matR_{\tt clu}$.
Importantly, $\matR_{\tt iso}$ is rank-deficient, particularly for below half-wavelength spacing because it depends on the aperture area and not the number of antennas.


\subsection{Mutual Coupling}

As the antenna elements at the BS are positioned in close proximity to each other, mutual coupling between elements causes distortion of the individual element's radiation patterns, which in turn affects the channel vector between the BS and the user. This effect can be introduced in the MIMO system model in \eqref{eq:rayleigh_channel} by multiplying the channel $\vech$ by the coupling matrix at the BS ${\matC \in \mathbb{R}^{M\times M}}$ as follows \cite[Eq. (105)]{circuit_theory_comm}
\begin{align}\label{eq:channel-model-coupling}
    \vech_{\tt mc} =  \matC^{1/2}\vech.
\end{align}

The combined effect of the mutual coupling in the array can be seen as a distortion of the channels observed by the BS, where some spatial dimensions are amplified and others are attenuated. From a mathematical viewpoint, the coupling matrix transforms the original coupling-free spatial correlation matrix $\matR$ into the effective spatial correlation matrix 
\vspace{-0.002\textheight}
\begin{align}
    \matR_{\tt mc} = \matC^{1/2} \matR \matC^{1/2}
\end{align}
of the effective channel $\vech_{\tt mc}$. Since $\matC$ is deterministic at the time scale in which the small-scale fading realizations occur, the effective channel also features correlated Rayleigh fading but with the new spatial correlation matrix $\matR_{\tt mc}$.

The coupling matrix of an antenna array, accounting for internal antenna losses, can be expressed as \cite{LucaHolo2024}
\vspace{-0.002\textheight}
\begin{align}
    \matC = \left(\matZ + {R_d}\matI \right)^{-1},
\end{align}
where $R_d>0$ represents the dissipation resistance of each antenna and $\matZ$ is the mutual impedance matrix. For a half-wavelength dipole the expression of
the dissipation resistance can be found in \cite[Example 2.13]{Balanis}. While this expression is applicable to any antenna array, obtaining analytical expressions for the mutual impedance matrix and antenna impedance is often challenging for many types of antenna elements. 
However, in the case of dipoles, which is the case in this paper, a more detailed derivation of the received voltages of an array of thin finite dipoles incorporating coupling effects can be found in \cite[Chapter 8]{Balanis}. More specifically, closed-form expressions for the mutual impedance can be found in \cite[Eqs. (8.69) and (8.71a-b)]{Balanis} for dipoles in side-by-side configuration. Closed-form expressions are also provided for dipoles in co-linear configuration\cite[Eqs. (8.72ab)]{Balanis}, and in parallel-in-echelon configuration \cite[Eqs. (8.73a-b)]{Balanis}.

\section{Channel Estimation}

It is essential in MIMO systems to have knowledge of the channel between each transmit-receive antenna link to implement downlink precoding and uplink combining schemes.
In time-division duplex (TDD) systems, the channel coefficients remain the same in the uplink and the downlink within the channel coherence time. In such scenarios, the standard procedure involves the user transmitting a predefined pilot sequence and the BS using the received signal to estimate the channel. Under mutual coupling and spatial correlation at the BS, the received signal $\vecy\in \mathbb{C}^M$ for the uplink pilot symbol transmission can be expressed as
\begin{align}
    \vecy = \sqrt{\rho}\vech_{\tt mc} + \vecn,
\end{align}
where $\rho>0$ is the pilot SNR and $\vecn$ is the zero-mean, unit-variance complex additive white Gaussian noise (AWGN).

An arbitrary linear estimator of  the channel $\vech_{\tt mc}$ based on the received signal $\vecy$ can be expressed as
\begin{align}\label{eq:estimated-channel}
    \widehat{\vech}_{\tt mc} = \matW\vecy,
\end{align}
where $\matW\in \mathbb{C}^{M\times M}$ is the determinstic matrix that characterizes the estimator.
For any $\matW$, the estimation error vector is $\vece\triangleq \vech_{\tt mc}-\widehat{\vech}_{\tt mc}$ and its covariance matrix is given by
\begin{align}
\begin{split}\label{eq:general-mse}
    \matR_{\tt e} = \matW\matR_{\tt y}\matW^{H} - \matR_{\tt h_{mc}y}\matW^{H} + {\matR_{\tt mc} - \matW\matR_{\tt h_{mc}y}^H},
\end{split}            
\end{align}
where $\matR_{\tt y}=\mathbb{E}\{ \vecy\vecy^H \}$ and $\matR_{\tt mc}=\mathbb{E}\{ \vech_{\tt mc}\vech_{\tt mc}^H \}$ are the auto-covariance matrices of  $\vecy$ and $\vech_{\tt mc}$, respectively, while $\matR_{\tt h_{mc}y}=\mathbb{E}\{ \vech_{\tt mc}\vecy^H \}$ is the cross-covariance matrix between the vectors $\vech_{\tt mc}$ and $\vecy$.

\subsection{MMSE-Structured Estimator}

Since the channel vector is complex Gaussian distributed, we adopt an estimator structured similarly to that of the MMSE estimator of complex Gaussian variables \cite{Kay1993SP}, expressed as:
\begin{align}\label{eq:mmse-structured-estimator}
\matW = \sqrt{\rho}\widehat\matR\left( \rho \widehat\matR + \matI \right)^{-1},
\end{align}
where $\widehat\matR$ represents an abstract covariance matrix. 
If $\widehat\matR = \matR_{\tt mc}$, we obtain the true (linear) MMSE estimator that minimizes the mean-squared error (MSE) (i.e., the trace of the error covariance matrix in \eqref{eq:general-mse}) among all choices of $\matW $.
The use of this estimator clearly requires knowledge of $\matR_{\tt mc}$.

However, practical scenarios often involve incomplete knowledge of channel statistics. The covariance matrix of the channel $\matR_{\tt mc}$ contains $M^2$ elements, making it challenging to acquire in holographic MIMO scenarios where $M$ is large. In particular, when the user transmits a small data packet or moves over large areas, it is not possible to observe sufficiently many independent channel realizations to make the sample covariance matrix estimate close to the ideal channel statistics.
To address the lack of knowing $\matR_{\tt mc}$ in practice, we can instead utilize the MMSE estimator's structure in \eqref{eq:mmse-structured-estimator} with a different correlation matrix $\widehat\matR$ representing the general covariance matrix due to the BS array geometry, rather than specific user propagation environments. 

With the next proposition, we highlight three special cases of the covariance matrix $\widehat\matR$ in \eqref{eq:mmse-structured-estimator} and how they affect the column space $\mathcal{C}\left(\cdot\right)$ spanned by the plausible channel estimates.

\begin{proposition}\label{prop:estimation-subspaces}
    Consider a channel realization $\vech_{\tt mc}\in \mathbb{C}^{M}$ as described in \eqref{eq:channel-model-coupling} and let $\widehat{\vech}_{\tt mc}$ denote an estimate of $\vech_{\tt mc}$ according to \eqref{eq:mmse-structured-estimator}. Then, for the following three scenarios of $\widehat\matR$, the column space of the channel estimates is as follows:
    \begin{enumerate}
        \item $\widehat\matR = 
        \matC^{\frac{1}{2}} \matR^{\frac{1}{2}} \matC^{\frac{1}{2}}
        \hspace{1.8mm}\Rightarrow \widehat{\vech}_{\tt mc} \in \mathcal{C}\left( \matC^{\frac{1}{2}} \matR^{\frac{1}{2}} \right)$,
        \item $\widehat\matR = 
        \matC^{\frac{1}{2}} \matR_{\tt iso}^{\frac{1}{2}} \matC^{\frac{1}{2}}
        \Rightarrow \widehat{\vech}_{\tt mc} \in \mathcal{C}\left( \matC^{\frac{1}{2}} \matR_{\tt iso}^{\frac{1}{2}} \right)$,
        \item $\widehat\matR = \matR_{\tt iso}
        \hspace{10.4mm}\Rightarrow \widehat{\vech}_{\tt mc} \in \mathcal{C}\left( \matR_{\tt iso}^{\frac{1}{2}} \right)$,
    \end{enumerate}
    where it holds that $\mathcal{C}\left( \matC^{\frac{1}{2}}\matR^{\frac{1}{2}} \right) \subseteq \mathcal{C}\left( \matC^{\frac{1}{2}}\matR_{\tt iso}^{\frac{1}{2}} \right)$.
    \begin{IEEEproof}
        See Appendix B.
    \end{IEEEproof}
\end{proposition}

Proposition \ref{prop:estimation-subspaces} provides valuable insights into the performance of the MMSE-structured estimator outlined in \eqref{eq:mmse-structured-estimator}, depending on the selection of the covariance matrix.
In the first scenario, we employ the true MMSE estimator, where the channel estimate $\widehat{\vech}_{\tt mc}$ is obtained by projecting the observed signal $\vecy$ onto the subspace spanned by the column vectors of $\matC^{\frac{1}{2}} \matR^{\frac{1}{2}}$. It's worth noting that in the absence of additive noise (e.g., $\rho\rightarrow \infty$), the observed signal $\vecy$ lies within the subspace of the channel, resulting in zero estimation error.

The second case occurs when the BS lacks precise knowledge regarding the user's channel-specific spatial correlation $\matR$ but has information about the mutual coupling, which can be measured beforehand or over an extended period of time. By setting $\widehat\matR = \matC^{1/2}\matR_{\tt iso}\matC^{1/2}$, it ensures consideration of all physically possible channel dimensions by considering an isotropic scattering environment where all dimensions are present. Any potential channel exists in the corresponding subspace, but for a particular spatial correlation matrix, only part of the subspace is utilized.

In the third scenario, the BS lacks knowledge of both the mutual coupling and spatial correlation of the user's channel. Hence, it assumes an isotropic scattering environment and disregards mutual coupling among antennas by setting $\widehat\matR = \matR_{\tt iso}$. In this case, the channel estimates lie in the subspace spanned by the column vectors of $\matR_{\tt iso}$, which may differ from the subspace of the actual channel. Such discrepancies can result in significant errors in the channel estimates.

\subsection{Least Squares Estimator}

When both the mutual coupling matrix and the spatial channel covariance matrix or the array geometry are unknown at the BS, an alternative classical approach is to use the LS estimator, which only requires knowledge of the pilot SNR, $\rho$. It is a linear estimator as in \eqref{eq:estimated-channel} obtained with
\begin{align}\label{eq:ls-estimator}
    \matW_{\tt ls} = \frac{1}{\sqrt{\rho}}\matI_M.
\end{align}
The LS estimator uses a special form of a projection matrix, where the channel estimate $\widehat\vech_{\tt mc}$ is obtained by projecting the observed signal $\vecy$ onto the entire vector space $\mathbb{C}^M$. 

\subsection{Mean Squared Error Analysis}

Now, let the eigenbases of $\matR_{\tt mc}$ and $\matW$ be denoted by $\matU_{\tt h_{mc}}$ and $\matU_{\tt w}$, respectively, and consist of the eigenvectors denoted by $\vecu_{\tt h_{mc},m}$ and $\vecu_{\tt w,n}$, respectively. Additionally, we define the eigenvalues $\lambda_{\tt h_{mc},m}$ and $\lambda_{\tt w,n}$ associated with the eigenvectors $\vecu_{\tt h_{mc},m}$ and $\vecu_{\tt w,n}$, respectively. 
In the next proposition, we express the MSE of the channel estimates in terms of the eigenbases and eigenvalues of the channel statistics and an arbitrary linear estimator.
\begin{proposition}\label{prop:mse-eigenvalues-eigenbases}
    Consider a channel realization $\vech_{\tt mc}\in \mathbb{C}^{M}$ and let $\widehat{\vech}_{\tt mc}$ denote an arbitrary linear estimator of $\vech_{\tt mc}$ according to \eqref{eq:estimated-channel}. Then, the MSE of the estimator can be expressed as
    \begin{align}\label{eq:proposition-mse-equation}
            {\tt MSE}   = \sum_{k=1}^M\sum_{l=1}^M \beta_{k,l} |\langle \vecu_{{\tt w},k} , \vecu_{{\tt h_{mc}}, l} \rangle|^2            
            + \sum_{l=1}^M \lambda_{{\tt h_{mc}},l},
    \end{align}
    where the scalar $\beta_{k,l}$ is given by 
    \begin{align}
        \beta_{k,l} = \left( \rho \lambda_{{\tt h_{mc}},l} + 1 \right) \lambda_{{\tt w},k}^2 - 2\sqrt{\rho}\lambda_{{\tt h_{mc}},l}\lambda_{{\tt w},k}
    \end{align}
    and $\langle \vecu_{{\tt w},k} , \vecu_{{\tt h_{mc}}, l} \rangle$ denotes the inner product between the eigenvectors $\vecu_{{\tt w},k}$  and $\vecu_{{\tt h_{mc}}, l}$.
    \begin{IEEEproof}
       This result can be derived by making an eigen-decomposition of the expressions in \eqref{eq:general-mse}.
    \end{IEEEproof}
\end{proposition}

Proposition \ref{prop:mse-eigenvalues-eigenbases} brings important insights into the relationship between the MSE of the channel estimate and the underlying eigenbases and eigenvalues of the estimator and channel statistics. For example, we can consider the MMSE-structured estimator in \eqref{eq:mmse-structured-estimator} with another correlation matrix that does not represent a particular user but the general array geometry. Then, its MSE is given by the following corollary.
\begin{corollary} \label{cor:mmse-arbitrary-correlation}
    Consider a channel realization $\vech_{\tt mc}\in \mathbb{C}^{M}$ and let $\widehat{\vech}_{\tt mc}$ denote an estimate of $\vech_{\tt mc}$ obtained using the estimation structure of \eqref{eq:mmse-structured-estimator} with an arbitrary covariance matrix $\widehat\matR$. Then, the MSE of the estimator can be expressed as in \eqref{eq:proposition-mse-equation} with
    \begin{align}
        \beta_{k,l} =  \frac{ \lambda_{{\tt h_{mc}},l} + \frac{1}{\rho} }{\left( \lambda_{{\tt w},k} + \frac{1}{\rho} \right)^2}\lambda_{{\tt w},k}^2 - \frac{2\lambda_{{\tt h_{mc}},l}\lambda_{{\tt w},k}}{ \lambda_{{\tt w},k} + \frac{1}{\rho} }.
    \end{align}
\end{corollary}
Corollary \ref{cor:mmse-arbitrary-correlation} highlights that for finite values of SNR, there is a penalty incurred when the correlation matrix of the estimator and that of the channel are not aligned in the same eigenbasis. However, as the SNR increases without bound (i.e., $\rho\rightarrow\infty$), this penalty diminishes, resulting in zero MSE. In this special case, where we employ the MMSE estimator, the eigenbasis of the channel and that of the estimator are common, i.e., $\vecu_{{\tt h_{mc}}, k} = \vecu_{{\tt w}, k}$ while  $\lambda_{{\tt h_{mc}},k} = \lambda_{{\tt w},k}$, $\forall k$. Then, the MSE becomes $\text{MSE} = \sum_{k=1}^M \lambda_{{\tt h_{mc}}, k}\left(\rho\lambda_{{\tt h_{mc}}, k} + 1 \right)^{-1}$. Comparing the latter expression with that in Corollary \ref{cor:mmse-arbitrary-correlation}, we see that the penalty can be high when the correlation matrix of the estimator and that of the channel have different eigenbases.

Notice that the eigenvalues of the LS estimator are $ \lambda_{{\tt w},k}=1/\sqrt{\rho}$, $\forall k$ while its eigenbasis is the same as that of the correlation matrix of the channel. Therefore, from Proposition \ref{prop:mse-eigenvalues-eigenbases}, we have that $\text{MSE}_{\tt ls} = M/\rho$\cite{Kay1993SP}.


\section{Numerical Results}

\begin{figure*}
\begin{multicols}{2}
  \includegraphics[width=\linewidth]{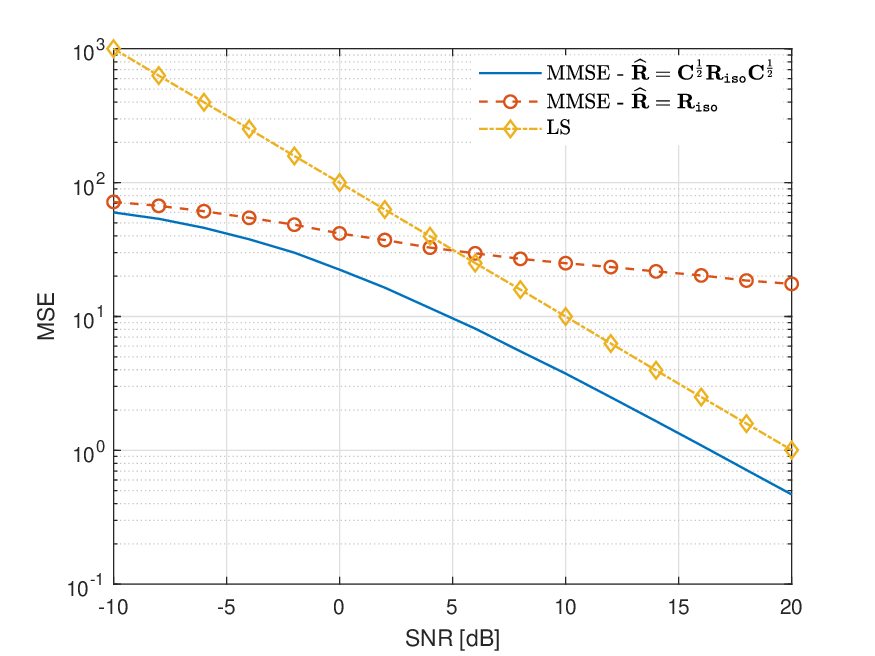}\caption*{(a) Isotropic scattering}
  \includegraphics[width=\linewidth]{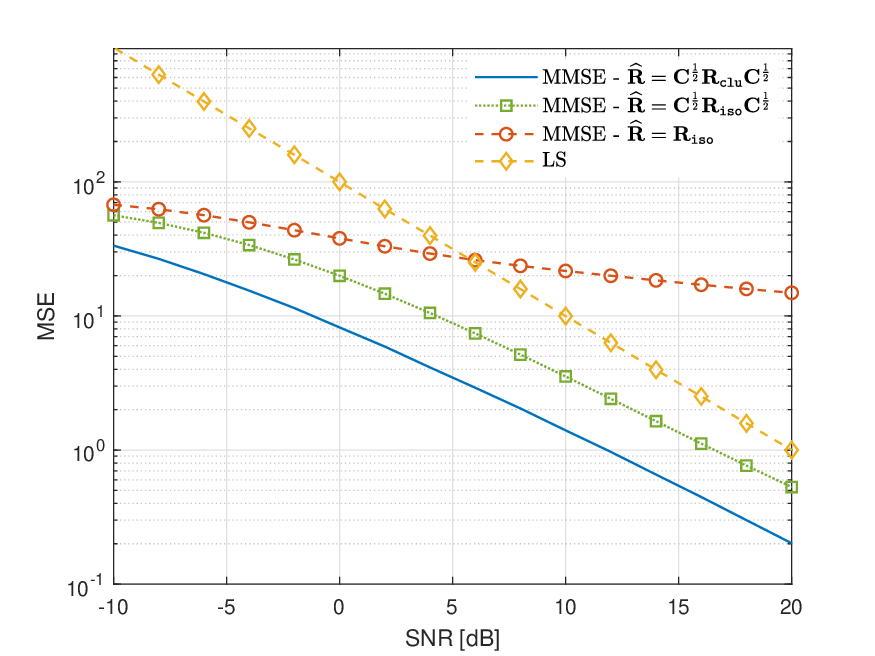}\caption*{(b) Non-isotropic scattering}
\end{multicols} \vspace{-3mm}
\caption{Channel estimates MSE as a function of the pilot SNR. In this example the antenna spacing is $\lambda/5$.} \label{figure:mse_vs_snr}
\end{figure*}

We will now quantitatively assess the channel estimation performance of the considered schemes by evaluating the MSE as a function of the pilot SNR, as depicted in Fig.~\ref{figure:mse_vs_snr}. Specifically, we investigate the performance of the estimator in \eqref{eq:mmse-structured-estimator} across various covariance matrices $\widehat\matR$, alongside the LS estimator in \eqref{eq:ls-estimator}, which operates independently of channel statistics.
Our analysis covers both isotropic and non-isotropic scattering environments, utilizing a $10\times 10$ UPA and an antenna spacing of $\lambda/5$.

\subsection{Isotropic Scattering Environment}

In the isotropic scattering environment, the spatial correlation matrix $\matR_{\tt iso}$ of the channel is calculated based on \eqref{eq:isotropic-correlation-formula}. Then, considering the mutual coupling effect, the covariance matrix of the channel $\matR_{\tt mc} = \matC^{1/2} \matR_{\tt iso} \matC^{1/2}$.

In Fig.~\ref{figure:mse_vs_snr}(a), we show how the MSE varies with the SNR.
We observe that when the BS utilizes the structure of the MMSE estimator in \eqref{eq:mmse-structured-estimator} while ignoring the influence of mutual coupling (i.e., $\widehat\matR = \matR_{\tt iso}$), the performance degradation at low SNR levels is negligible compared to the true MMSE estimator. However, as the SNR increases, the performance deterioration becomes significant. For instance, at SNRs of $10$ and $20$\,dB, the performance loss is approximately $8$ and $16$\,dB, respectively.

Therefore, awareness of the mutual coupling matrix is essential for the MMSE estimator's performance. In cases where information about the mutual coupling is unavailable, the BS can opt for the LS estimator, which doesn't require knowledge of the mutual coupling or spatial channel correlation matrices. While the LS estimator exhibits considerable performance degradation at low SNR levels compared to the MMSE, while the performance gap narrows to $4$\,dB at higher SNR levels.

\subsection{Non-isotropic Scattering Environment}

The non-isotropic scattering environment is modeled as in \cite{ChanEstimOzlem2022}; that is, there are $N = 20$ clusters with exponential power delay profiles that are generated according to the urban macro-cell environment model in sub-6 GHz frequency band by following\cite[p. 54]{Series2017}. The BS and user are $25$\,m and $1.5$\,m above the ground, respectively. The nominal azimuth and elevation angles are determined from the cluster powers according to\cite[p. 55-58]{Series2017}. The per-cluster angular standard deviations are $\sigma_\phi = \sigma_\theta = 2^{\circ}$.

For the non-isotropic scattering environment, the spatial correlation matrix $\matR_{\tt clu}$ of the channel is calculated based on closed-form approximation in \cite[Lemma 2]{ChanEstimOzlem2022}. Then, considering the mutual coupling effect, the covariance matrix of the channel becomes $\matR_{\tt mc} = \matC^{1/2} \matR_{\tt clu} \matC^{1/2}$.

We first consider the scenario where the BS employs an estimator with the structure of MMSE but lacks awareness of both mutual coupling and the spatial correlation of the user's specific channel, i.e., $\widehat\matR = \matR_{\tt iso}$. In Fig.~\ref{figure:mse_vs_snr}(b), we observe that ignoring mutual coupling leads to $4$\,dB performance degradation at low SNR levels compared to the MMSE estimator. However, as the SNR increases, the performance deterioration becomes significant. For instance, at SNR values of $10$ and $20$\,dB, the performance loss is approximately $12$ and $19$\,dB, respectively, which is higher than the isotropic scattering scenario (cf. Fig.~\ref{figure:mse_vs_snr}(a)).

Next, if the BS employs an estimator with the MMSE-structure and has perfect knowledge of the mutual coupling but is unaware of the spatial correlation of the user's specific channel, i.e., $\widehat\matR = \matC^{1/2}\matR_{\tt iso}\matC^{1/2}$, the performance loss compared to the true MMSE estimator is approximately $4$\,dB across the entire SNR range.

Therefore, using isotropic spatial correlation instead of the user's specific channel correlation reduces the MMSE estimator's performance by $4$dB. We also conclude that awareness of the mutual coupling matrix is more critical than knowing the spatial correlation of the channel for MMSE estimator performance.
Finally, the LS estimator exhibits considerable performance degradation at low SNR levels compared to the true MMSE estimate, while the performance reduces to $7$dB at higher SNR levels.



\section{Conclusion}

In this paper we analyzed the impact of mutual coupling and spatial correlation information on channel estimation precision in holographic MIMO systems. We considered several MMSE-structured estimators and compare them with the LS estimator. Specifically, when considering the MMSE estimator, we distinguished between two kinds of prior information: mutual coupling information and user-specific spatial channel correlation. We analyzed the impact when the BS has access to both, one or none of them. It was shown that it is crucial of considering mutual coupling when implementing channel estimation, while having access to the spatial channel correlation is less important. Finally, it was shown that the LS estimator is resilient to the absence of mutual coupling information but only performs well at high SNR.


\section*{Appendix A: Proof of Proposition \ref{prop:spatial-correlation-isotropic}}
The expression \eqref{eq:isotropic-correlation-formula} is derived by rewriting the exponential term $x_{nm}\triangleq \veck(\varphi,\theta)^T \left(\vecr_n - \vecr_m\right)=d^{nm}_y\sin \varphi \cos \theta + d^{nm}_z\sin \theta$ in the integral \eqref{eq:spatial-correlation-formula} according to Euler's formula and then expanding its sinusoidal terms using a Taylor series around zero as
\begin{align}
    \cos x_{nm} &= \sum_{k=0}^{\infty} \frac{(-1)^kx_{nm}^{2k}}{(2k)!}\label{eq:real_term},\\
    \sin x_{nm} &= \sum_{k=0}^{\infty} \frac{(-1)^kx_{nm}^{2k+1}}{(2k+1)!}\label{eq:imag_part},
\end{align}
where the $\cos$ and $\sin$ terms correspond to the real and imaginary parts of the exponential, respectively. Additionally, the horizontal and vertical distances between antenna $m$ and $n$ (normalized by the wavelength) are given by $d_y^{nm} = {\left(r_y(n) - r_y(m)  \right)d_y}/{\lambda}$ and $d_z^{nm} ={\left(r_z(n) - r_z(m)  \right)d_z}/{\lambda}$, respectively.
Then, using the binomial theorem to expand the term $x_{nm}^{2k}$ in \eqref{eq:real_term}, the real part of the integral \eqref{eq:spatial-correlation-formula} becomes
\begin{align*}
\begin{split}
        [\matR_{\tt iso}]_{nm}=\sum_{k=0}^{\infty} \sum_{l=0}^{k} \frac{(-1)^k}{(2k)!}\binom{2k}{2l}\left( d_z^{nm} \right)^{2(k-l)} \left(d_y^{nm} \right)^{2l}\\
        \times \iint \limits_{-\pi/2}^{-\pi/2} \sin^{2(k-l)}\theta \cos^{2(l+2)}\theta \sin^{2l}\varphi .
\end{split}
\end{align*}
We then use the following integrals for the power of trigonometric functions 
\cite[TI (232) and TI (226)]{tableOfIntegrals2007} to obtain  the closed-form expressions:
\begin{align*}
\begin{split}
    \int_{-\pi/2}^{\pi/2} \sin^{2l} xdx = \frac{1}{2^{2l}} \binom{2l}{l}\pi
\end{split}
\\[1ex]
\begin{split}
    \int_{-\pi/2}^{\pi/2} \sin^{2(k-l)}x \cos^{2(l+2)}xdx = \frac{1}{2^{2(k-l)}} \binom{2(k-l)}{k-l}\pi \\
    \times \frac{(2l+3)!!}{(2k+4)(2k+2)...(2k-2l+2)}
\end{split}
\end{align*}

Following similar procedure, we can find that the imaginary part in \eqref{eq:imag_part} is zero\cite[TI (227)]{tableOfIntegrals2007}.

\section*{Appendix B: Proof of Proposition \ref{prop:estimation-subspaces}}
We notice from \eqref{eq:estimated-channel} that the channel estimate $\widehat\vech_{\tt mc}$ lies in the column space of the channel estimator. In the first scenario, where $\widehat\matR = \matR_{\tt mc}$, the estimator in \eqref{eq:mmse-structured-estimator} can be rewritten as 
\begin{align}
    \matW = \matU_{\tt h_{mc}} \underbrace{\frac{1}{\sqrt{\rho}}\boldsymbol{\Lambda}_{\tt h_{mc}} \left( \boldsymbol{\Lambda}_{\tt h_{mc}}+\frac{1}{\rho}\matI  \right)^{-1}}_{\triangleq \boldsymbol{\Lambda}_{\tt w}}   \matU_{\tt h_{mc}}^H,
\end{align}
where the matrix $\matU_{\tt h_{mc}}$ and the diagonal matrix $\boldsymbol{\Lambda}_{\tt h_{mc}}$  denote the eigenbasis and its corresponding eigenvalues of $\matR_{\tt mc}$, respectively.
Then, the column space of the estimator can be determined from its eigenvalue decomposition. For a channel of rank $r$, we have
\begin{align*}
    \mathcal{C}\left( \matW \right) &= \{\matW\vecy: \vecy \in \mathbb{C}^M \}\\
       &= \{\matU_{\tt h_{mc}}\boldsymbol{\Lambda}_{\tt w}\underbrace{\matU_{\tt h_{mc}}^H\vecy}_{\triangleq\vecx}: \vecy \in \mathbb{C}^M \}\\
       &=\{\matU_{\tt h_{mc}}\boldsymbol{\Lambda}_{\tt w}\vecx: \vecx \in \mathbb{C}^M \}\\
       &= \left\{ \sum_{k=1}^r x_k\lambda_{\tt w,k}\vecu_{\tt h_{mc},k}  \right\}\\
       &= \text{span}\left\{\vecu_{\tt h_{mc},1}, \vecu_{\tt h_{mc},2}\dots \vecu_{\tt h_{mc},r} \right\}.
\end{align*}
Thus, the column space of the estimator is spanned by the columns of $\matU_{\tt h_{mc}}$ which are the eigenvectors of the matrix $\matR_{\tt mc} = \matC^{\frac{1}{2}}\matR\matC^{\frac{1}{2}} = \matC^{\frac{1}{2}}\matR^{\frac{1}{2}} ( \matC^{\frac{1}{2}}\matR^{\frac{1}{2}} )^T$. Notice that $\matU_{\tt h_{mc}}$ also gives the orthonormal basis for the column space of $\matC^{\frac{1}{2}}\matR^{\frac{1}{2}}$ \cite[p. 367]{Strang2006Linear} and thus $\mathcal{C}\left( \matW \right) = \mathcal{C}( \matC^{\frac{1}{2}}\matR^{\frac{1}{2}} )$. Similarly, it can be proved in the second and third scenarios. Finally, we can prove that $\mathcal{C}\left( \matC^{\frac{1}{2}}\matR^{\frac{1}{2}} \right) \subseteq \mathcal{C}\left( \matC^{\frac{1}{2}}\matR_{\tt iso}^{\frac{1}{2}} \right)$ using fact that the subspace spanned by the columns of $\matR_{\tt iso}$ contains the subspace spanned by any other $\matR$.





\bibliographystyle{IEEEtran}
\bibliography{bibtex/IEEEabrv,bibtex/confs-jrnls,bibtex/publishers,bibtex/referbib}

\end{document}